# Deep Bayesian Local Crystallography

Sergei V. Kalinin,[1,a] Mark P. Oxley,[1] Mani Valleti,[2] Junjie Zhang,[3,b] Raphael P. Hermann,[3] Hong Zheng[4], Wenrui Zhang[1,3c], Gyula Eres[3], Rama K. Vasudevan,[1] Maxim Ziatdinov[1,5]

[1] Center for Nanophase Materials Sciences, Oak Ridge National Laboratory, Oak Ridge, TN 37831

[2] Bredesen Center for Interdisciplinary Research, The University of Tennessee, Knoxville, TN 37996

[3] Materials Science and Technology Division, Oak Ridge National Laboratory, Oak Ridge TN 37831

[4] Materials Science Division, Argonne National Laboratory, Argonne, IL 60439, USA

[5] Computational Sciences and Engineering Division, Oak Ridge National Laboratory, Oak Ridge TN 37831

The advent of high-resolution electron and scanning probe microscopy imaging has opened the floodgates for acquiring atomically resolved images of bulk materials, 2D materials, and surfaces. This plethora of data contains an immense volume of information on materials structures, structural distortions, and physical functionalities. Harnessing this knowledge regarding local physical phenomena necessitates the development of the mathematical frameworks for extraction of relevant information. However, the analysis of atomically resolved images is often based on the adaptation of concepts from macroscopic physics, notably translational and point group symmetries and symmetry lowering phenomena. Here, we explore the bottom-up definition of structural units and symmetry in atomically resolved data using a Bayesian framework. We demonstrate the need for a Bayesian definition of symmetry using a simple toy model and demonstrate how this definition can be extended to the experimental data using deep learning networks in a Bayesian setting, namely rotationally invariant variational autoencoders.

[a] Sergei2@ornl.gov
[b] Current affiliation: State Key Laboratory of Crystal Materials & Institute of Crystal Materials, Shandong University, Jinan, Shandong 250100 China
[c] Current affiliation: Ningbo Institute of Materials Technology and Engineering, Chinese Academy of Sciences, Ningbo, Zhejiang, 315201 China

Macroscopic symmetry is one of the central concepts in the modern condensed matter physics and materials science. Formalized via point and spatial group theory, symmetry underpins areas such as structural analysis, serves as the basis for the descriptive formalism of quasiparticles and elementary excitations, phase transitions, and mesoscopic order-parameter-based descriptions, especially of crystalline solids. In macroscopic physics, symmetry concepts arrived with the advent of X-ray methods developed by Bragg, and for almost a century remained the primary and natural language of physics. Notably, the rapid propagation of laboratory X-ray diffractometers and large-scale X-ray scattering facilities provided ample experimental data across multiple material classes and serve as a necessary counterpart for theoretical developments. Correspondingly, symmetry-based descriptors have emerged as a foundational element of condensed matter physics and materials science alike.

The natural counterpart of symmetry-based descriptors is the concept of physical building blocks. Thus, crystalline solids can be generally described via a combination of the unit cells with discrete translational lattice symmetries. At the same time, systems such as Penrose structures possess well-defined building blocks but undefined translation symmetry. Finally, a broad range of materials lack translational symmetries, with examples ranging from structural glasses and polymers to ferroelectric and magnetic morphotropic systems.[1-9] Remarkably, the amenability of symmetry-based descriptors have led to much deeper insights into the structure and functionalities of materials with translational symmetries compared to (partially) disordered systems.[10-12]

The beginning of the 21st century has seen the emergence of real space imaging methods including scanning probe microscopy (SPM)[13-15] and especially (scanning) transmission electron microscopy ((S)TEM).[16-18] Following the introduction of the aberration corrector in the late '90s[19] and the advent of commercial aberration-corrected microscopes, atomically resolved imaging is now mainstream. Notably, modern STEMs allow atomic columns to be imaged with ~pm-level precision.[20] This level of structural information allows insight into the chemical and physical functionalities of materials, including chemical reactivity, magnetic, and dielectric properties utilizing structure-property correlations developed by condensed matter physicists from macroscopic scattering data.[21-27] Over the last decade, several groups have extended these analyses to derive mesoscopic order parameter fields such as polarization,[28-31] strains and chemical strains,[32] and octahedra tilts[33-35] directly from STEM and SPM data. In several cases, these data can be matched to the mesoscopic Ginzburg-Landau models, providing insight into the generative mesoscopic physics of the material.[36, 37] Recently, a similar approach was proposed and implemented for theory-experiment matching via microscopic degrees of freedom.[38-40]

Yet, despite the wealth of information contained in atomically resolved imaging data, analyses to date were almost invariably based on the mathematical apparatus developed for macroscopic scattering data. However, the nature of microscopic measurements is fundamentally different. For the case of ideal single crystal containing a macroscopic number of structural units, the symmetry of the diffraction pattern represents that of the lattice and the width of the peaks in Fourier space is determined by the intrinsic factors such as angle resolution of the measurement system, rather than disorder in the material. The presence of symmetry breaking distortions, such as the transition from a cubic to tetragonal state, is instantly detectable from peak splitting. For microscopic observations only a small part of the object is visible and the positions of the atoms are known only within an uncertainty interval; this uncertainty can be comparable to the magnitude of the symmetry breaking feature of interest such as tetragonality or polarization. Thus, questions arise: What image size is it justified to define the symmetry from the atomically resolved data? and What level of confidence can be defined? Ideally, such an approach should be applicable not only for structural data, but also for more complex multi-dimensional data sets such as those available in scanning tunneling spectroscopy (STS)[41] in scanning tunneling microscopy (STM), force-distance curve imaging[42] in atomic force microscopy, or electron energy loss spectroscopy (EELS)[43, 44] and ptychographic imaging[45-47] in scanning transmission electron microscopy (STEM).

Here we propose an approach for the analysis of spatially resolved data based on deep learning in a Bayesian setting. This analysis utilizes the synergy of three fundamental concepts; the (postulated) parsimony of the atomic-level descriptors corresponding to stable atomic configurations, the presence of distortions in the idealized descriptors (e.g., due to local strains or other forms of symmetry breaking), and the presence of possible discrete or continuous rotational symmetries. These concepts are implemented in a workflow combining feature selection (atom finding), a rotationally invariant variational autoencoder to determine symmetry invariant building blocks, and a conditional autoencoder to explore intra-class variability via relevant disentangled representations. This approach is demonstrated for 2D imaging data but can also be generalized for more complex multi-dimensional data sets.

## 1. Why local symmetry is Bayesian

Here, we illustrate why the consistent definitions of local symmetry properties necessitates the Bayesian framework. As an elementary, but easy to generalize example, we consider the 1D diatomic chain formed by alternating atoms (1) and (2) with coordinates generated by the rule $x_i^{(1)} = x_i^{(2)} + a$, $x_{i+1}^{(2)} = x_i^{(1)} + b$. The atomic coordinates with some uncertainty stemming from the observational noise, sampling, etc. are experimentally observed and hence, the atomic positions, $x_j^{exp}$, that are the sum of the ideal positions, $x_j^{(1,2)}$, and noise, $\delta$, are available for observation. We assume that the atom types are not observed (e.g. they have similar contrast), i.e., atoms (1) and (2) are indistinguishable. Correspondingly, we aim to answer the question - what number of observations can distinguish the simple chain, $a = b$, and diatomic chain, $a \neq b$? Note that this problem is equivalent to, e.g., distinguishing a square and tetragonal unit cell and can be generalized to more complex cases with the addition of several parameters.

The classical answer to this question is given by frequency-based statistics. Here, an alternative hypothesis (i.e. single vs. double chain) is formed where the point estimates for the average lattice parameters and their dispersions are calculated and the $p$-test can be used determine the correctness of the hypothesis. However, this approach has several significant limitations: it does not consider any potential prior knowledge of the system, it implicitly relies on the relevant distributions being Gaussian, and it is sensitive to the choice of an ideal system. A detailed analysis of the relevant drawbacks is given by Kruschke.[48]

An alternative approach to these problems is via the Bayesian framework, based on the concept of prior and posterior probabilities linked as:[49, 50]

$$p(\theta_i|D) = \frac{p(D|\theta_i)p(\theta_i)}{p(D)} \qquad (1)$$

where $D$ represents the data obtained during the experiment, $p(D|\theta_i)$ represents the likelihood that this data can be generated by the model, $i$, with parameter, $\theta_i$. The prior, $p(\theta_i)$, reflects the prior knowledge about the model. The posterior, $p(\theta_i|D)$, describes the new knowledge (i.e., updated model and model parameters) as a result of the observational data. Finally, $p(D)$ is the denominator that defines the total space of possible outcomes.

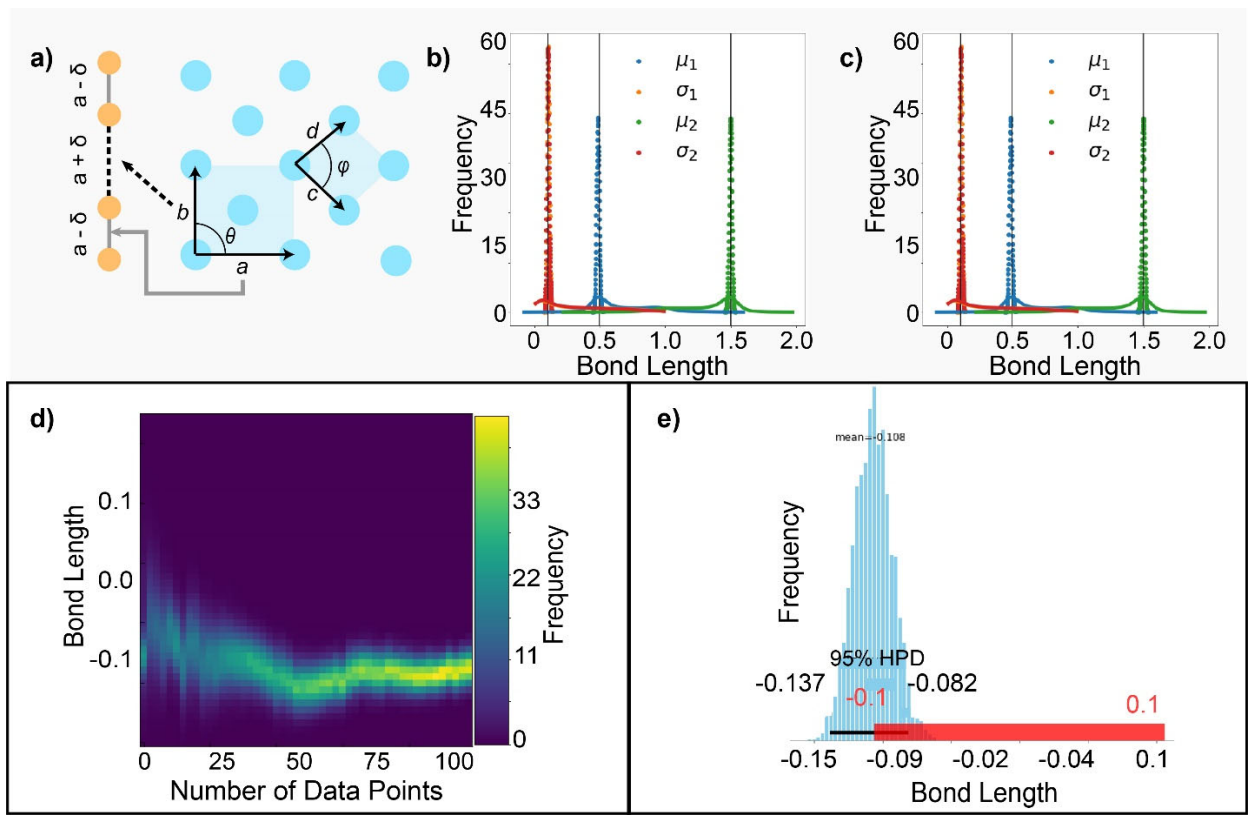

**Figure 1.** (a) Schematic illustration of the correspondence between even-odd chains and 2D lattice. (b) Final posterior distributions of parameters ($\mu_1$, $\mu_2$, $\sigma_1$, and $\sigma_2$) involved in analysis of case – 1, posterior distributions after first update are shown by solid lines. (c) Final posterior distributions of parameters ($\mu_1$, $\mu_2$, $\sigma_1$, and $\sigma_2$) involved in case – 2, posterior distributions after first update are shown by solid lines. (d) Posterior distribution of $\mu_3$ as a function of number of datapoints for case -2. (e) High density interval (HDI) of $\mu_3$ (blue) and region of practical importance (ROPE) (red).

As an example, a set of diatomic chains is generated with bond lengths derived from two normal distributions, $N(\mu = 0.5, \sigma^2 = 0.01)$ and $N(\mu = 1.5, \sigma^2 = 0.01)$, where $\mu$ is the men and $\sigma$ the standard deviation of the distribution. These two sets of bond lengths are treated independently and are referred to as odd and even bond lengths, respectively. The likely distributions for this case are also assumed to be normal distributions, $N(\mu = \mu_1, \sigma = \sigma_1)$ and $N(\mu = \mu_2, \sigma = \sigma_2)$. A total of four parameters, $\mu_1$ and $\sigma_1$ for the odd bond lengths and $\mu_2$ and $\sigma_2$ for the even bond lengths, exhaustively determine the parameter space. We refer to this analysis as case – 1.

The key element of Bayesian inference is the concept of prior, summarizing the known information on the system.[49-51] In experiments, the priors are typically formed semi-quantitatively based on general physical knowledge of the material (e.g., $SrTiO_3$ is known to be cubic with lattice parameter 3.1Å). For this model example, the prior distributions of all four parameters are formed based on the first 10 observations, $Y_{10}$. The prior distribution for $\mu_1$ and $\mu_2$ is a Laplace distribution, $L(\mu = Y_{10}, b = 0.2*Y_{10})$, whereas for $\sigma_1$ and $\sigma_2$ it is a uniform distribution, $U(0, Y_{10})$. This method of prior selection removes any *a priori* bias about the sample and only uses data obtained from experimental images. However, the priors can also be obtained from known materials properties (assuming a perfect imaging system). The posterior distributions of the parameters are updated with each datapoint. Fig. 1a shows a schematic of how the odd and even chain analyses can be extended to a more general 2D Bravais lattice. Fig. 1b shows the final posterior distributions of the parameters involved, with the posterior distributions after an update with first respective datapoints of each set shown by the solid lines. The means of both normal distributions are close to the real values and are far away from each other.

For a non-trivial case, odd and even bond lengths are derived from the normal distributions, $N(\mu = 0.95, \sigma^2 = 0.01)$ and $N(\mu = 1.05, \sigma^2 = 0.01)$. Here, the difference in the means is on the order of the standard deviation. We refer to this analysis as case – 2. Fig. 1b shows the final posterior distributions of the parameters involved, with the posterior distributions after the first update shown by the solid lines. To answer the question of whether the set of bond lengths belong to a simple chain or a diatomic chain, we construct the distribution of the difference in bond lengths with a likelihood, $N(\mu = \mu_3, \sigma = \sigma_3)$. For a simple ideal lattice, this distribution should be centered at zero with no standard deviation. Fig. 1d shows the posterior distribution for $\mu_3$ as a function of the number of datapoints. We then construct an interval region of practical equivalence (ROPE), which is the region around the hypothesis where the hypothesis is still true. A decision on the validity of the hypothesis can be made by comparing the highest density interval (HDI, 94% credible interval) and the ROPE. For illustration purposes, the ROPE is considered to be [-0.1, 0.1] in Fig. 1e and the HDI for $\mu_3$ is also shown. Decision rules for different overlaps of HDI and ROPE are discussed in e.g., Ref. [49].

This simple example illustrates that for microscopic observations, many fundamental parameters are defined only in the Bayesian sense as the posterior probability densities. They can be related to the macroscopic definitions though concepts such as the practical equivalence. For large system sizes, the Bayesian estimates converge to the macroscopic model. We pose that the Bayesian descriptions of symmetry and structural properties from the bottom up should be Bayesian in nature, updating the prior knowledge of the system with the experimental data.

## 2. Local crystallographic analysis

As a second concept, we discuss established approaches for the systematic analysis of atomic structures from experimental observations and the deep fundamental connections between the intrinsic symmetries present (or postulated) in the data and the neural network architectures. For example, the classical fully connected multilayer perceptron intrinsically assumes the presence of potential strong correlations between arbitrarily separated pixels of the input image, resulting in a well-understood limitation of these networks to only relatively low-dimensional features. Convolutional neural networks (CNNs) are introduced as a universal approach for equivariant data analysis where the features of interest can be present anywhere within the image plane. This network architecture implicitly assumes the presence of continuous translational symmetry, similar to the sliding window/transform approach, Figure 2 (a).[52-54] While allowing derivation of mesoscopic information, even for atomically resolved data, this approach suffers from inevitable spatial averaging and ignores the existence of well-defined atomic units.

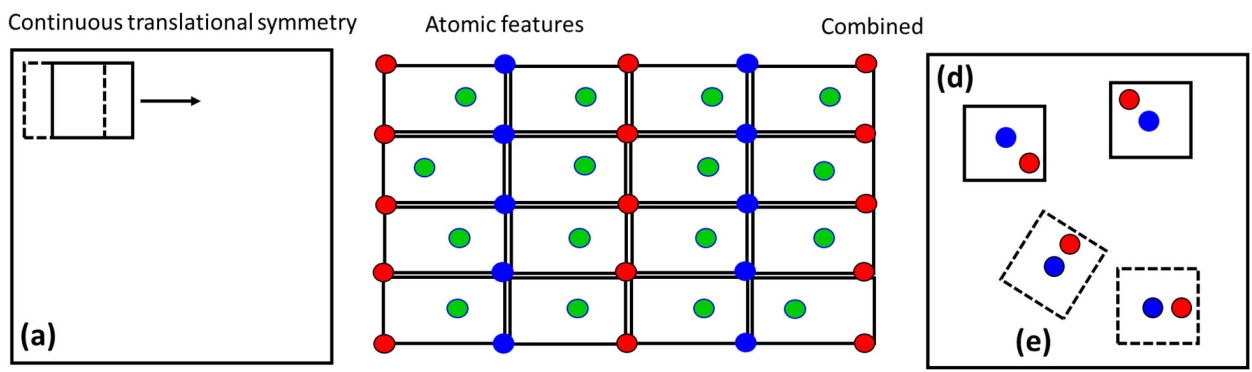

**Figure 2.** Local structural analysis for systems with (a) continuous translational symmetry, (b) known discrete translational symmetry, and (b) discrete system without translational symmetry and unknown local symmetries.

If the positions of the atomic species can be determined, the analysis can be performed based on the local atomic neighborhoods (local crystallography)[55, 56] or the full atomic connectivity graph. In these approaches, the full image is reduced to atomic coordinates and the subsequent analysis is based on the latter. It is important to note that in this case all remaining information in the image plane is ignored, i.e., the full data set is approximated by the point estimates of the atomic positions. Finally, the combined approach can be based on the analysis of sub-images centered on defined atomic positions.[57, 58] In this case, the known atomic positions provide the reference points and the sub-images contain information on the structure and functionality around them.

For atomic and sub-image-based descriptors, the behavior referenced to the ideal behavior is of interest and is defined by high-symmetry positions or ideal lattice sites. If these are known, then behaviors such as symmetry-breaking distortions can be immediately quantified and explored, Figure 2 (b). However, the very nature of experimental observations is such that this ground truth information is not available directly, necessitating suitable approximations. For example, an ideal lattice can be postulated and average parameters can be found using a suitable filtering method. However, this approach is sensitive to minute distortions of the image (e.g., due to drift) and image distortion correction is required. Similarly, variability in the observed images due to microscope configurations (mistilt, etc.) can provide observational biases.

These examples illustrate that deep analysis of the structure and symmetry from atomically resolved data sets necessitates simultaneous (a) identification of ideal building blocks and symmetry breaking distortions, while (b) allowing for general rotational invariance in the image plane and (c) accounting for discrete translational symmetry as implemented in the Bayesian setting. Ideally, such descriptors will be referenced to local features, as shown in Figure 2 (c).

## 3. Bayesian local crystallography

Here, we aim to combine the local crystallography and Bayesian approaches. The general workflow for deep Bayesian local crystallographic analysis is shown in Figure 3 (a). For the first step, the STEM image or a stack of images are fed into the deep fully convolutional neural network (DCNN). for semantic segmentation and atom finding.[59, 60] The semantics segmentation refers to a process where each pixel in the raw experimental data is categorized as belonging to an atom (or to a particular type of atom) or to a "background" (vacuum). The atom finding procedure is then performed on segmented data by finding a center of the mass of each segmented blob (corresponding to an atom) with a sub-pixel precision. The details of the DCNN configuration and implementation are available via the AtomAI repository on GitHub.[61] The DCNN-derived atomic positions are used to define the stack of sub-images centered on each atom and represent the image contrast in the vicinity of each atom.

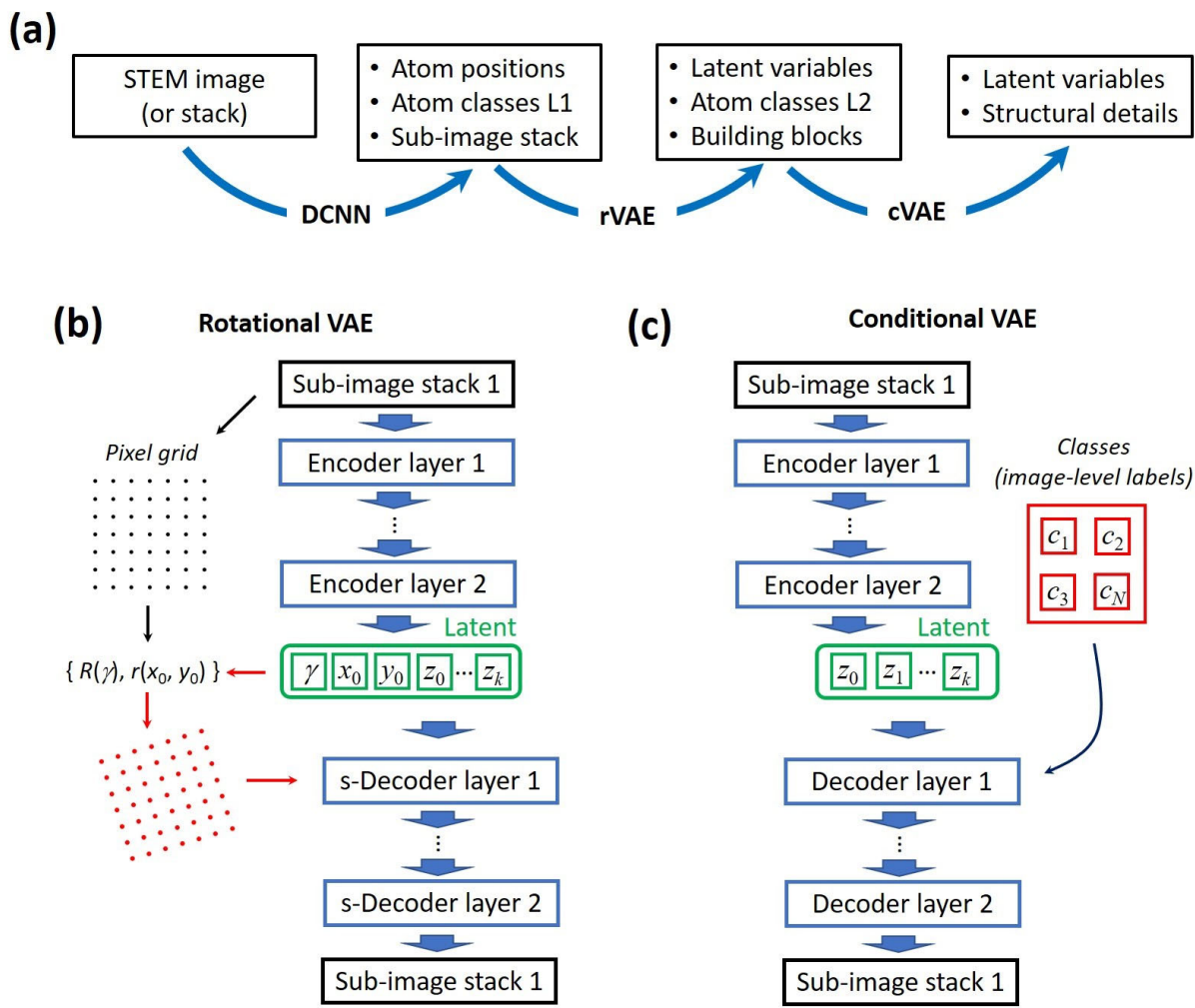

**Figure 3.** (a) General workflow of deep Bayesian local crystallographic analysis. (b) Schematic of rotational variational autoencoder (rVAE) workflow. (c) Schematic of conditional variational autoencoder (cVAE) workflow. Note that for the sake of simplicity of the schematics the VAE's reparameterization trick[62] is not shown. In rVAE, the encoder layers can be either fully-connected or convolutional, whereas the decoder layers are always fully-connected layers (in the current implementation, inclusion of convolutional layers breaks the rotational symmetry). In cVAE, layers in encoder and decoder can be either fully connected or convolutional depending on the problem.

Note that this sub-image description is chosen since both the original STEM data and DCNN reconstructions contain information beyond atomic coordinates, such as column shapes and unresolved features, and this needs to be taken into account during analysis. It is important to note that the choice of sub-image stack (original image, smoothed image, or DCNN output) defines the type of information that will be explored. For example, DCNN outputs define the probability density that a certain image pixel belongs to a given atom class that is optimal for exploration of chemical transformation pathways. At the same time, original image contrast may be optimal for exploration of physical phenomena. Finally, we note that the extremely important issue in this analysis is the correction of distortions for effects such as fly-back delays or general image instabilities, which can alleviate unwanted artifacts and introduce new ones. Several examples of these will be discussed below. If necessary, these sub-images can be used to further refine the classes using standard methods such as principal component analysis (PCA) or Gaussian mixture modelling (GMM). However, as mentioned above, these clustering methods will tend to separate the atoms into symmetry equivalent positions, leading to over-classification and poorly separable classes.

To avoid this problem, the subsequent step in the analysis is the rotationally invariant variational autoencoder (rVAE). In general, VAE is a directed latent-variable probabilistic graphical model. It allows learning stochastic mapping between an observed $x$-space (in this case, space of the sub-images) with a complicated empirical distribution and a latent $z$-space whose distribution can be relatively simple. The VAE consists of generative and inference models, which are Bayesian networks of the form $p(x|z)p(z)$ and $q(z|x)$, respectively. For the generative model, the latent variable, $z_i$, is a "code" (hidden representation) from which it reconstructs $x_i$. The potentially complex, non-linear dependency between $x_i$ and $z_i$ is parameterized by a (deep) neural network (NN) with weights $\theta$, $p_\theta(x|z)$, which takes "code" $z_i$ as an input. The inference model is used to approximate the posterior of the generative model, $p_\theta(z|x)$, and represents a flexible family of variational distributions parameterized by NN with weights $\phi$ and $q_\phi(z|x)$. The NN-parameterized inference and generative models are frequently referred to as encoder and decoder, respectively. The parameters of the two networks ($\theta$ and $\phi$) are jointly learned by maximizing the evidence lower boundary (ELBO) consisting of the reconstruction loss term and Kullback-Leibler divergence term with a mini-batch stochastic gradient descent.

Here, we aim to learn a rotationally invariant code for our data. Unfortunately, standard neural network layers (fully connected and convolutional) do not respect rotational symmetry or invariance. One potential way to circumvent this problem is to use convolutional layers with modified, steerable filters.[63] Another approach, which is specific to the VAE set up, is to disentangle rotations and translations from image content by making the generative model (decoder) explicitly dependent on the coordinates (Figure 3b).[64] In this case, the ELBO is computed as

$$ELBO = Reconstruction\ Loss - D_{KL}\big(q(z|x)\|\mathcal{N}(0,I)\big) - D_{KL}\big(q(\gamma|x)\|\mathcal{N}(0,s_\gamma^2)\big), \quad (2)$$

where $\gamma$ is a latent angle (see Figure 3b) and $s_\gamma$ is a "rotational prior" set by a user before the optimization. The second and third terms in Eq. (2) are the KL divergences associated with image content and rotation angle, respectively.

Hence, for the rVAE, the latent space is configured to comprise the rotational angle, the two variables allowing for the offset in $x$ and $y$ directions, and additional unstructured latent variables. This configuration is ideally suited for the analysis of the variability in the STEM sub-image stack since small uncertainties in the atomic positions are naturally accommodated through the offset latent variables and the continuous or discrete rotations are captured by the angle variable. The remaining latent variables can be used in a manner similar to classical variational autoencoders to explore the variability of structural patterns in the latent space.

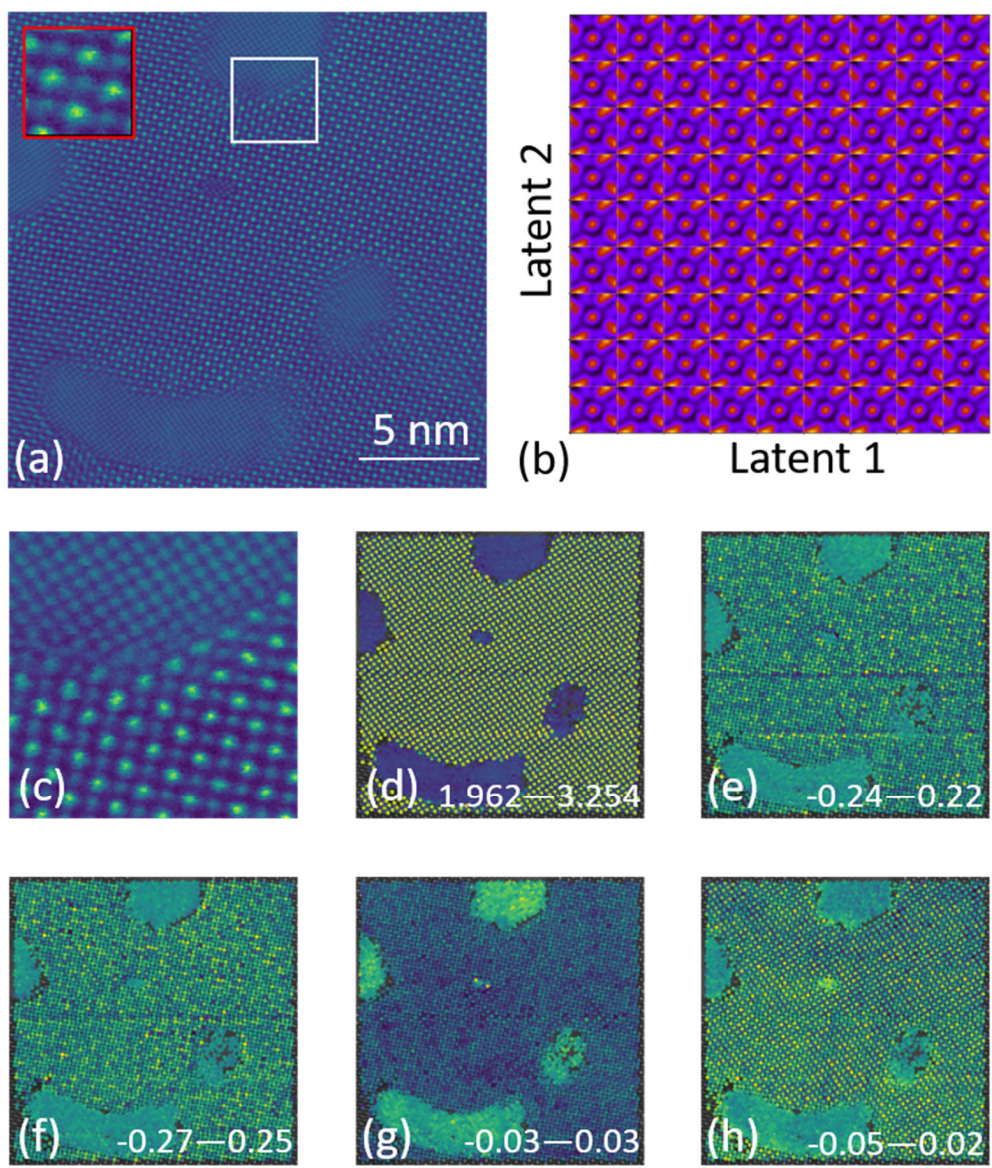

**Figure 4.** rVAE analysis of multiphase (La$_x$Sr$_{1-x}$)MnO$_3$ – NiO system using a window size of 36 pixels. Shown are (a) original image, (b) sub-image representation in 2D latent parameter space, (c) zoom-in of image in (a) showing phase morphology, (d) angle, (e,f) *x* and *y* offsets, and (g,h) latent variables 1 and 2. Analysis is performed on raw images. Insets in (d)-(f) show range of variation in image intensities.

The rVAE analysis of the multiphase system is illustrated in Figure 4. Here, Fig. 4 (a) shows the atomically resolved STEM image of the multiphase (La$_x$Sr$_{1-x}$)MnO$_3$ (LSMO) – NiO system. The dense NiO inclusions with a rock salt structure in the perovskite LSMO matrix are clearly observed, as visualized in Fig. 4 (c). DCNN allows one to locate virtually all the atomic units in the LSMO matrix and a majority of the atoms in the NiO. The sub-image stack formed from the DCNN output was analyzed by rVAE, Fig. 4 (b), is a representation of the atomic configurations in the latent space of the system. For the window size of 36 pixels used here, there is little variation in either the latent 1 or latent 2 directions.

The encoded angle, Fig. 4 (d), shows a clear checkerboard pattern in the LSMO phase and is uniform in the NiO phase. The offset maps shown in Fig. 4 (e,f) are relatively featureless but contain horizontal lines that are attributed to minute scanning non-idealities. The spatial maps of the latent components are shown in Fig. 4 (d-h). An interesting contrast behavior is observed in the latent space; latent parameter 1 shows a clear variation between the NiO and LSMO phases but little variation within each phase. Latent parameter 2, Fig. 4 (h), exhibits the checkerboard pattern of the LSMO perovskite lattice but is relatively featureless within the NiO phases. These results can be understood by examining the variable histograms (Supplementary Fig. S1). The encoded angle is clearly split between two peaks and the latent space histograms also indicate a separation of features. The offsets, however, form single peaks that account for the lack of strong features observed in Fig. 4 (e,f).

The effect of varying the window size is shown in Supplementary Fig. S2 where the analysis is repeated for window sizes of 24 and 48 pixels. For the smaller window, the histogram peaks of the encoded angle are broadened and the checkerboard pattern in the spatial map becomes less distinct. The histograms of the latent spaces split into two peaks and the spatial maps both reflect the checkerboard pattern in the LSMO perovskite phase. For completeness, we repeated the analysis on the DCNN segmented data (Supplementary Figures S3—S5) and it proved far more sensitive to window size; a 24--pixel window is used in the analysis shown in Fig. S3. The most notable difference compared to the raw analysis is a distinct transition from the perovskite to the rock--salt structure in the sub-image representation in the 2D latent parameter space in the latent variable 1 direction, as shown in Fig. S3 (b). The perovskite phase of the first latent variable is almost featureless but is strongly differentiated from the NiO phase. While the histogram of the encoded angle in Fig. S4 is split into two peaks, as shown in Fig. S5, increasing or decreasing the window size leads to the encoded angle collapsing into small variations about a single value.

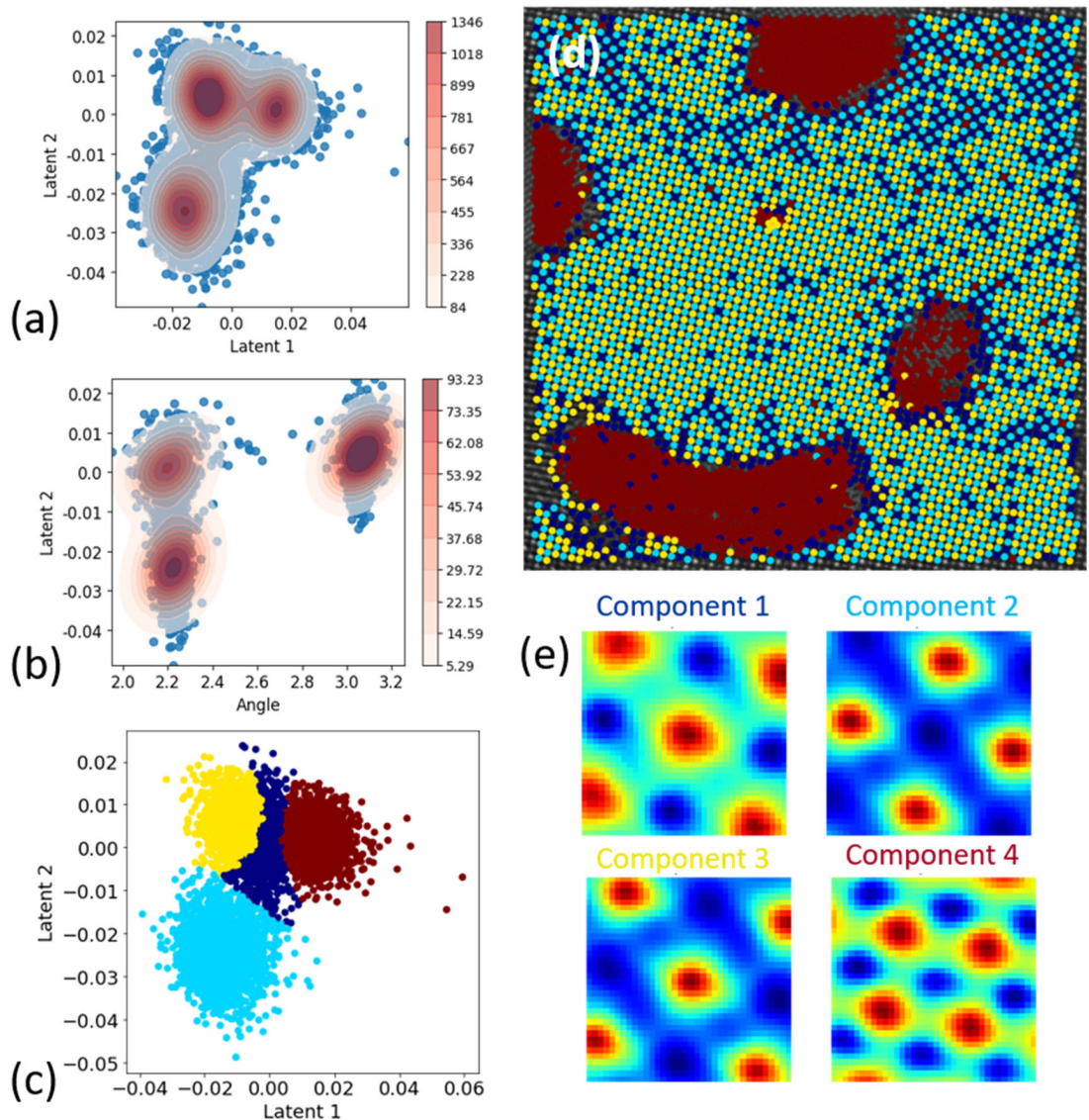

**Figure 5.** Latent space distributions. Pair distribution for (a) latent variables and (b) encoded angle and second latent variable. (c) Gaussian mixture model (GMM) clustering for latent spaces. (d) Original STEM image with superimposed class labels and GMM centroid images corresponding to four components used. (e) Four components: component 1 (dark blue), component 2 (cyan), component 3 (yellow), component 4 (brown). Analysis performed on raw data using a window size of 36 pixels.

The structure of the distributions in the latent space of rVAE was used to gain further insight into the analysis, as shown in Fig. 5. For the offsets (not shown), the distribution is close to Gaussian with most of the points localized within ~20% of the origin. In comparison, the latent space distribution exhibits extremely interesting behavior. Figure 5 (a) shows the joint distribution of the latent variables, visualized both as individual points and with a superimposed kernel density estimate (KDE). Note that each point corresponds to the sub-image and describes the behavior of the local neighborhood of a single lattice atom. The representation as points and KDE allows the comparison between the total system behavior (including the distribution of outliers) and the corresponding densities (average behaviors) and is necessary given the large number of points (from ~$10^4$ for single images to ~$10^5$ for the stacks).

To further extend this analysis, we note that the rVAE often tends to disentangle dissimilar types of distortions within a system. For example, experiments with a large number of different STEM images (beyond those shown in this paper) illustrate that scan distortions often tend to be described by one (group of) latent variables, whereas systematic changes in the local structure are described by the remaining latent variables. This property of VAEs is generally well known in computer science applications such as style networks; however, here we see that it applies for the physical systems as well.

We further explore this separation of atomic units based on neighborhood behavior using disentangled representations. As observed in Fig. 4, the angle and latent variable 2 seem to offer the optimal 2D basis to separate the atomic units, with clear contrast and a lack of distortion behaviors. The corresponding distribution and KDE plots are shown in Fig. 5 (b), illustrating three clearly defined groups of points corresponding to the A-site and B-site cations in the LSMO phase and columns in the NiO phases, respectively. Note that the KDE peaks corresponding to the three atomic types that jointly comprise >90% of points are fairly narrow. At the same time, there are a large number of outliers showing the presence of atoms with the behaviors falling on the continuous lines between the three groups, forming the manifold of possible states in the system.

Finally, we can gain further insight into the spatial distributions and classes of behaviors via clustering in the latent space. Fig. 5 (c) shows the Gaussian mixture model (GMM) clustering of points in the latent space. Note that given the complex structure of the distribution, the choice of a proper covariance matrix for the GMM, or the exploration of different clustering methods, will highlight different aspects of system behavior and hence offer a powerful tool for the exploration of corresponding physics. Here, we show as an example of the separation in three components. The spatial distribution of the label maps is shown in Fig. 5 (d) and images corresponding to the centroids of the GMM classes are shown in Fig. 4 (e). Components 3 (yellow) and 2 (cyan) correspond to the A and B sites in the perovskite, respectively, while component 4 (brown) corresponds to the NiO phase. Component 1 (dark blue) is a distorted form of component 3 and mostly clusters around the boundary with the NiO phase; it replaces component 3 elsewhere, suggesting the presence of local image distortions. The analysis is repeated on the DCNN segmented image shown in Fig. S6. In this case component 1 corresponds to the LSMO lattice and components 2 and 3 correspond to the NiO phase. Component 4 again clusters about the phase boundaries and occurs occasionally elsewhere in the LSMO phase.

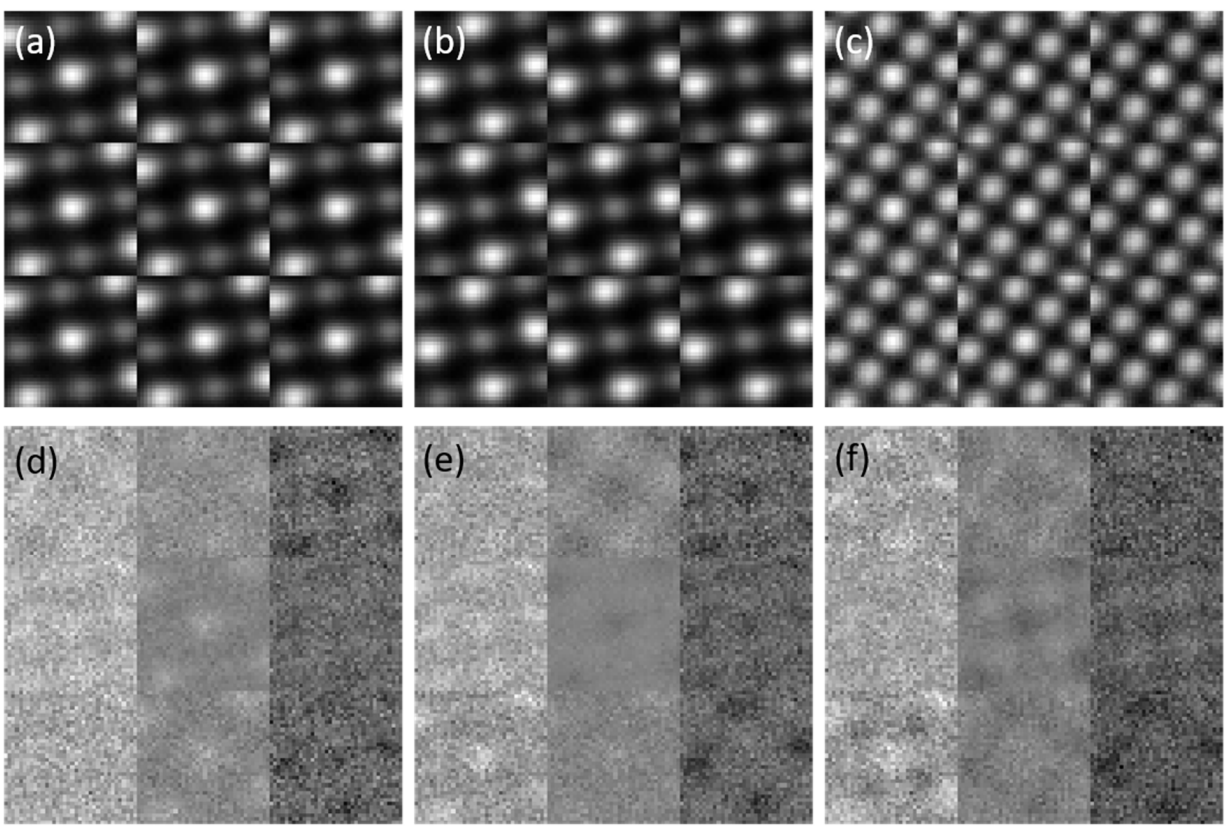

**Figure 6.** Conditional VAE analysis of the three GMM components form Fig. 5 corresponding to the (a,d) A-site (component 3), (b,e) B-site (component 2) columns of the LSMO phase and (c,f) NiO phase (component 4). The upper images are the raw images, and the lower images have the average of the 3x3 tableau subtracted.

To get further insight into the materials structure, we explore the disentangled representations of the structural building blocks using the conditional variational autoencoder (cVAE) approach. The schematics of cVAE is shown in Figure 3 (c). Here, the autoencoder approach is applied on the concatenated image stack (or its reduced representation) and the class labels. In this manner, the latent space encodes both the sub-images and labels. On the decoding stage, the reconstructed object is drawn from the combination of the latent variables and the label. The typical example of the cVAE application will be disentanglement of the styles in the MNIST data set. Whence simple VAE will draw all the numbers and distribute them in the latent space, the cVAE will draw the selected number and the latent space representations will reflect writing styles – e.g. tilt, line width, etc. The key aspect of using cVAE approach, as opposed of VAE analysis of individual classes, is that the thus disentangled styles will be common across the data set, reminiscent of hierarchical Bayesain models.

As an example of cVAE analysis, shown in Figure 6 is the latent space representation for the first three GMM components of the LSMO-NiO system. Here, the latent space is subdivided into 3x3 regions and the corresponding images are reconstructed. Shown are the images per se and the images with subtracted average. Note that while direct physical interpretation of this disentangled representations is complex, we note the commonality in the character of changes in the vertical and lateral directions for all three components. The central position of each tableau show the least variation from the mean value, with higher than average values seen to the left and lower than average values to the right.

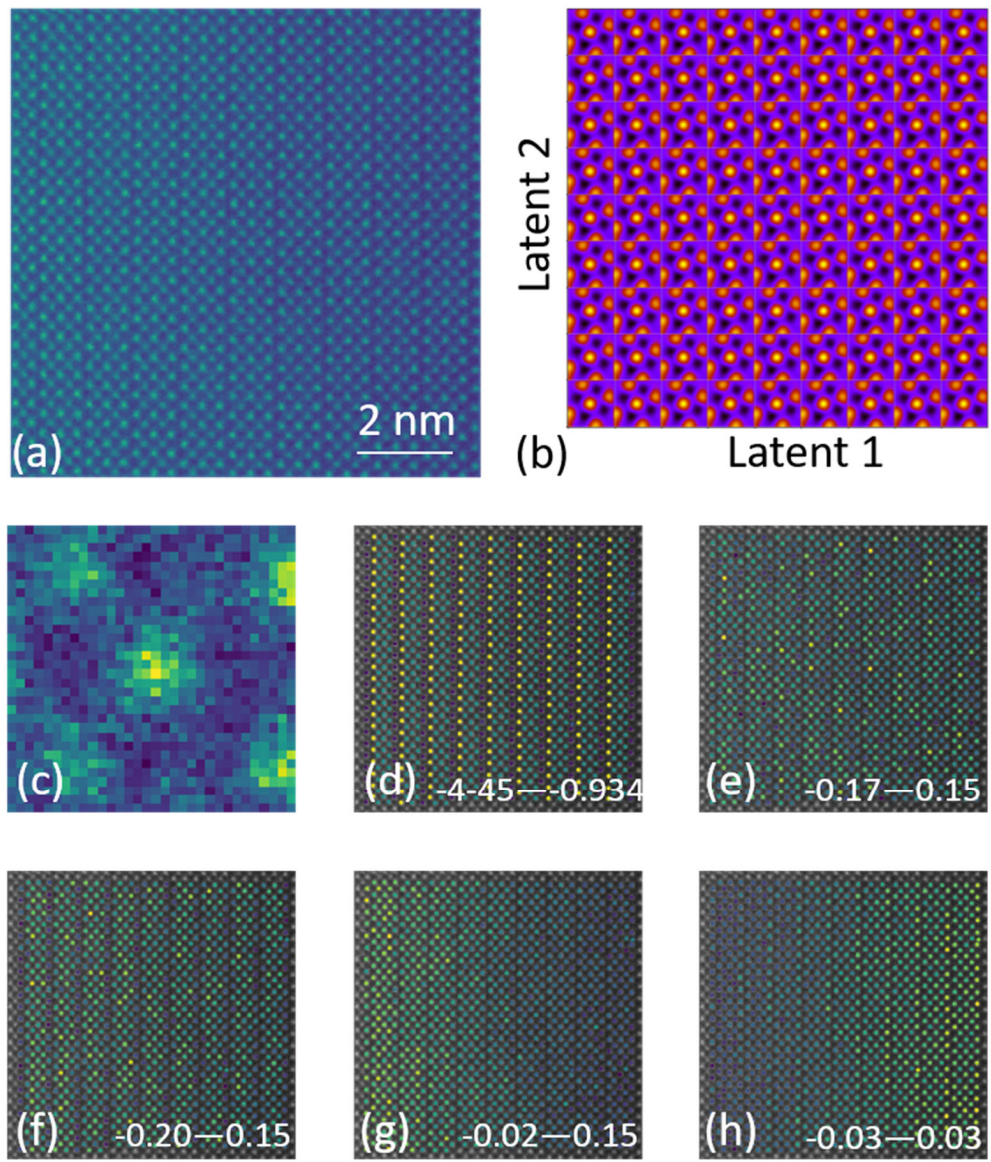

**Figure 7.** rVAE analysis of Sr₃Fe₂O₇ image. (a) Original STEM image, (b) sub-image representation in 2D latent parameter space, (c) a single sub-image used for analysis, (d) angle, (e,f) $x$ and $y$ offsets, and (g,h) latent variables 1 and 2. Analysis is performed on raw images using window size of 32 pixels. Insets indicate intensity variation of each panel.

We can extend this analysis to a system with a significantly more complex lattice such as the Sr₃Fe₂O₇ (SFO) layered perovskite. Sr₃Fe₂O₇ is a mixed valence Ruddlesden-Popper series compound with double perovskite structure that nominally features tetravalent iron. Charge disproportionation to Fe(III) and Fe(V) was observed by Mössbauer spectroscopy.[65, 66] Spiral magnetic order was observed by neutron diffraction[67] and provides a rare example of a magnetic cycloid arising from a ferromagnetic nearest neighbor competing with antiferromagnetic next-nearest exchange.[68] Further interest in this material arise from high oxygen mobility.[69] The preparation of a near stoichiometric compound requires high oxygen partial pressure.[70]

The rVAE analysis of SFO is shown in Figure 7. The original STEM image, Fig. 7 (a), clearly illustrates the layered structure of SFO. Of most interest is the encoded angle, Fig. 7 (d), which shows three separate values, one down the center of the layers and a different value on either edge of the layers. The histogram of the encoded angle is shown in Fig. S7 where three peaks are clearly present. The outer peaks differ by approximately $\pi$ radians, which is consistent with the edges of the layers experiencing a 180° rotation in their local configuration. The x-offset in Fig. 7 (e) is relatively featureless, which is consistent with the histogram shown in Fig. S7. The y-offset in Fig. 7 (f), however, clearly identifies the boundary between layer, with its histogram displaying two peaks. The two latent spaces in Fig. 7 (g-h) exhibit a gradual change in intensity from left to right corresponding to the sample thickness variation, which is also observed in the raw STEM image (Fig. 7 (a)). The corresponding histograms all have flattened peaks corresponding to this gradual change. Similar to observations for a 2-phase system, these behaviors are now disentangled and can be explored separately. We observed a similar separation for other STEM images where scan distortions e.g., due to fly-back delays, were clearly concentrated in a single latent variable.

The choice of window size is crucial for extracting some of these features. The effect of using a smaller and larger window size on the rVAE process is shown in Fig. S8. For a smaller window size, all the histograms exhibit single peaks with a restricted angular range. The latent spaces, however, still reflect thickness variations in the sample. For a larger window size, the encoded angle histogram exhibits only two peaks and only a single value is observed at the boundary of the layer. For completeness, this analysis was also performed on the DCNN segmented images and the results are shown in Fig. S8-S9. The behavior of the encoded angles shown in Fig. S8 is similar to that shown in Fig. 7. However, the latent variables in Fig. 7 do not reflect the thickness variation with the first latent variable in Fig. S8 (g) and in fact is quite featureless and the second latent variable strongly highlights the boundary between layers. Unlike the raw STEM image results, through the use of a smaller window size on the segmented data, the histogram of the encoded angle retains three peaks, as observed in Fig. S10. The histogram for the second latent variable is clearly split between two values and this is reflected in the corresponding spatial representation. For the larger window size, the encoded angle collapses to two peaks as before but both latent variables strongly show the layered structure.

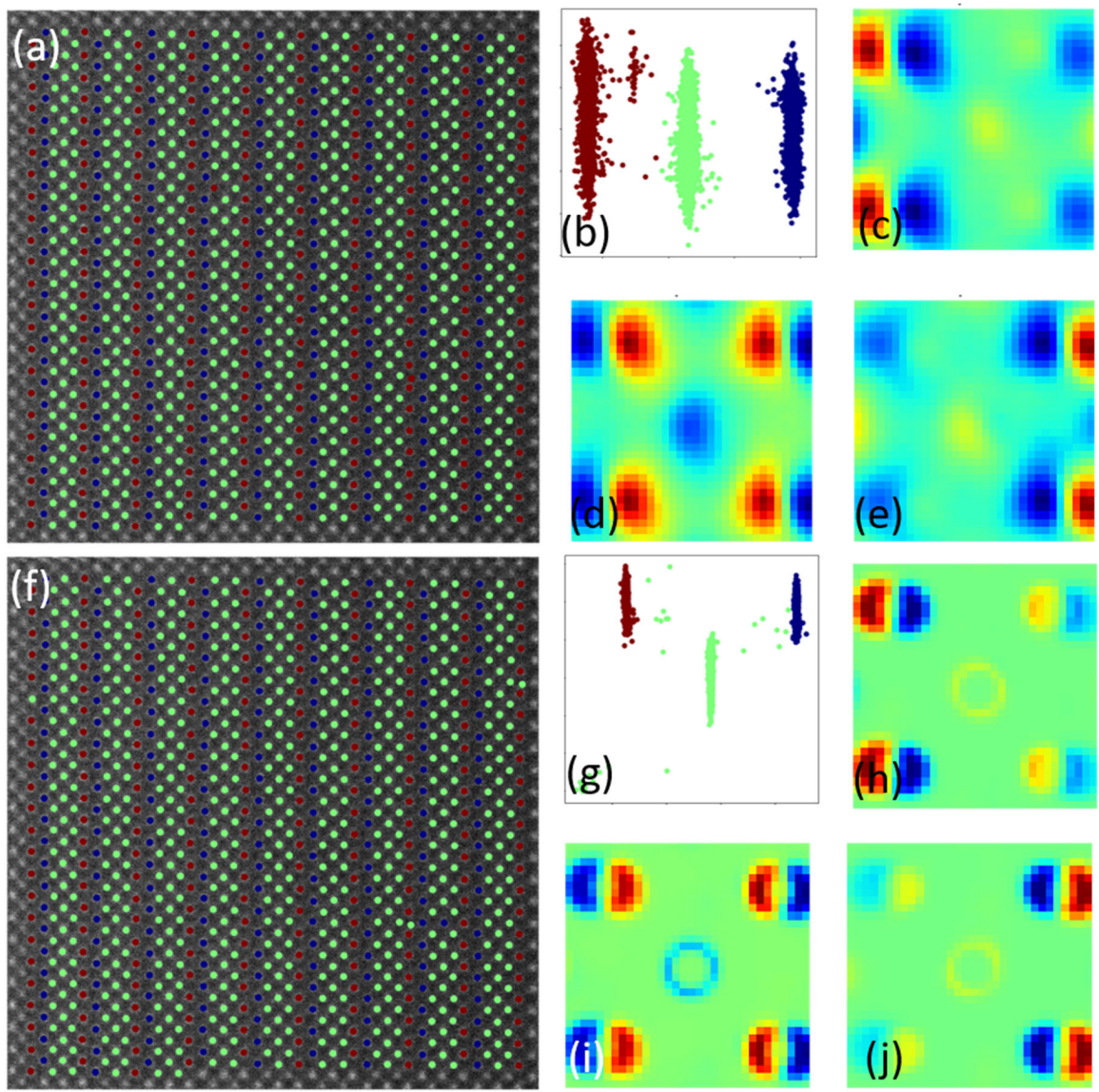

**Figure 8.** Analysis of SFO as a raw data (a-e) and DCNN semantically segmented (f-j) representations. Shown are (a,f) images with superimposed GMM markers, (b,g) GMM clusters in latent spaces, and (c-e) and (h-j) reconstructed components corresponding to GMM centroids.

Component 1 is represented by dark blue dots, with components 2 and 3 represented by green and brown dots, respectively. A window size of 32 pixels is used in both cases.

Clustering analyses of the latent representations for SFO are shown in Figure 8. Fig. 8 (a) shows the label map for the GMM clusters shown in Fig. 8 (b) where the clustering is performed over angles and the second latent variable. Note the presence of three well-separated atomic groups that are consistent with the three peaks in the encoded angle histograms previously discussed. Examination of components 1 and 3, shown in Fig 8 (c) and (e), respectively, show that they are near mirror images of each other. This is consistent with a 180° difference in the corresponding peaks in the histograms of the encoded angles. The second component, shown in Fig. 8 (d), is approximately centrosymmetric and corresponds to the center of the layered structure.

We compare the analysis of the raw images with the DCNN outputs. As mentioned above, the raw data represents the true variability of the STEM image contrast but also includes a high noise level. Comparatively, DCNN output is the semantically segmented image, i.e., the probability density that a specific image pixel belongs to the atom. Fig. 8 (f) shows the label map for the GMM clusters shown in Fig. 8 (g) where the clustering is performed over angles and the second latent variable. Once again there are three distinct clusters (although with some scatter for the second component). As with the raw image analysis, the first and third components are near mirror images and the second component is centrosymmetric.

Finally, we examine the effect of increasing the window size to 48 pixels in Fig. S11. For both the raw and segmented images the corresponding histograms of the encoded angle only exhibit two distinct peaks. Fig. S11 (a) shows the label map for the GMM clusters shown in Fig. S11 (b). For the raw image the components are closely clustered about two encoded angles, with some scatter noted for the third component. Rather than being mirror images, components 1 and 2 seem to exhibit inverse contrast. Component 3 is a distorted version of component 2 and mainly replaces component 2 on the label map. It does, however, occasionally replace component 1, which is expected due to the small cluster coinciding with the cluster of component 1 observed in Fig S11 (b). For the DCNN segmented data shown in Fig S11 (f)-(j), the result appears much clearer. Components 2 and 3 are reversed in contrast and are tightly clustered. Component 1 is randomly distributed in both Fig S11 (f) and (g).

To summarize, we introduce a workflow for the bottom-up symmetry and structural analysis of atomically resolved STEM imaging data. For systems with known or ad hoc defined rotational variants, the combination of Gaussian mixture modeling and principal component analysis (GMM-PCA) allows separation of the relevant structural units and structural distortions for individual units. However, the GMM-PCA combination fails in the presence of multiple rotational variants and especially general rotations, since in this case the class will be assigned to each rotation of the same structural unit. The use of the rVAE-cVAE approach proposed here allows one to generalize the classification-distortion analysis for the general rotational symmetry. We illustrate that the capability of VAEs to produce disentangled representations can be beneficially used to separate structural units, relevant distortions, and in certain cases, the instrumental distortions, opening the pathway for systematic studies of symmetry breaking distortions for a broad range of material systems.

While implemented here for the analysis of structural STEM images, we expect that a similar approach can be used for the analysis of symmetry breaking distortions in e.g., scanning tunneling microscopy (STM) images, and can be further extended to the analysis of multidimensional data sets such as tunneling spectroscopy in STM, EELS and 4D STEM in STEM, and so on. Furthermore, similar to other Bayesian methods, it will be of interest to explore physics-based prior distributions in the latent space, beyond the class labels used here. Overall, we believe that the combination of the capability to disentangle physical phenomena via latent space representations and parsimonious analysis makes the proposed workflow universal for multiple physical problems.

**Acknowledgements:**

This effort (ML, STEM, film growth, sample growth) is based upon work supported by the U.S. Department of Energy (DOE), Office of Science, Basic Energy Sciences (BES), Materials Sciences and Engineering Division (S.V.K., S.V., G.E., W.Z., J.Z., H.Z., R.P.H.) and was performed and partially supported (R.K.V., M.Z.) at the Oak Ridge National Laboratory's Center for Nanophase Materials Sciences (CNMS), a U.S. Department of Energy, Office of Science User Facility. Dr. Matthew Chisholm is gratefully acknowledged for the STEM data used in this work. Dr. Katharine Page is gratefully acknowledged for help in the data acquisition at NOMAD. A portion of this research used resources at the Spallation Neutron Source, a DOE Office of Science

User Facility operated by the Oak Ridge National Laboratory. The authors are deeply grateful to Dr. Karren More for careful reading and correcting the manuscript.

**Materials and methods:**

**Thin film growth.** The LSMO-NiO VAN and the single-phase LSMO and NiO films were grown on STO(001) single-crystal substrates by PLD using a KrF excimer laser ($\lambda$= 248 nm) with fluence of 2 J/cm$^2$ and a repetition rate of 5 Hz. All films were grown at 200 mTorr $O_2$ and 700 °C. The films were post-annealed in 200 Torr of $O_2$ at 700 °C to ensure full oxidation, and cooled down to room temperature at a cooling rate of 20 °C/min. For out-of-plane transport measurements, the films were grown on 0.5% Nb-doped STO(001) single-crystal substrates. The film composition was varied by using composite laser ablation targets with different composition.

**Sample preparation**

A polycrystalline rod of $Sr_3Fe_2O_{7-x}$ with 6 mm in diameter and 50 mm in length was prepared using powders synthesized from solid state reaction of stoichiometric $SrCO_3$ and $Fe_2O_3$ at 1100 °C. The single crystalline material utilized here was grown using a high pressure floating zone furnace with $O_2$ partial pressure of 148 bar. Refinement of neutron diffraction data obtained at the NOMAD instrument of the Spallation Neutron Source using GSAS-II[71] revealed a single-phase material with an oxygen content of 6.8, see Supplementary Fig. S12 and Table S1.

**STEM:**

The plan-view STEM samples of Ni-LSOM were prepared using ion milling after mechanical thinning and precision polishing. In brief, a thin film sample was firstly ground, and then dimpled and polished to a thickness less than 20 micrometer from the substrate side. The sample was then transferred to an ion milling chamber for further substrate-side thinning. The ion beam energy and milling angle were adjusted towards lower values during the thinning process, which was stopped when an open hole appeared for STEM characterization. The $Sr_3Fe_2O_7$ sample(s) were prepared by FIB lift out followed by local low energy Ar ion milling, down to 0.5 eV, in a Fischione NanoMill.

The STEM used for the characterization of both samples was a Nion UltraSTEM200 operated at 200 kV. The beam illumination half-angle was 30 mrad and the inner detector half-angle was 65 mrad. Electron energy-loss spectra were obtained with a collection half-angle of 48 mrad.

# Deep Bayesian Local Crystallography

Sergei V. Kalinin, Mark P. Oxley, Mani Valleti, Junjie Zhang, Raphael P. Hermann,

Hong Zheng, Wenrui Zhang, Gyula Eres, Rama K. Vasudevan, Maxim Ziatdinov

## Supplementary Figures

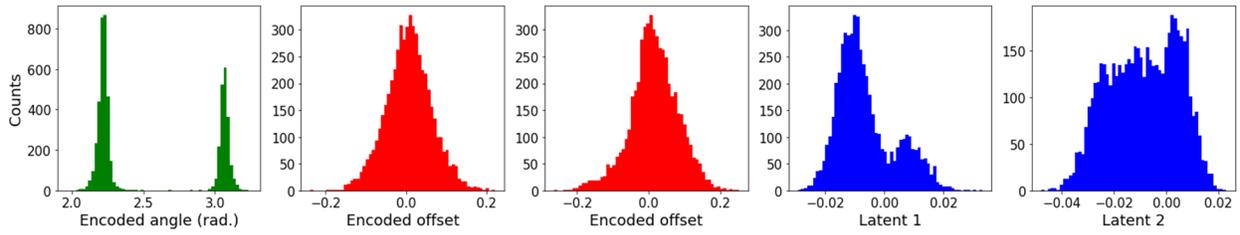

**Figure S1:** Histograms of encoded angle, offsets, and latent spaces for (La$_x$Sr$_{1-x}$)MnO$_3$ – NiO system shown in Fig. 4 using window size of 36 pixels.

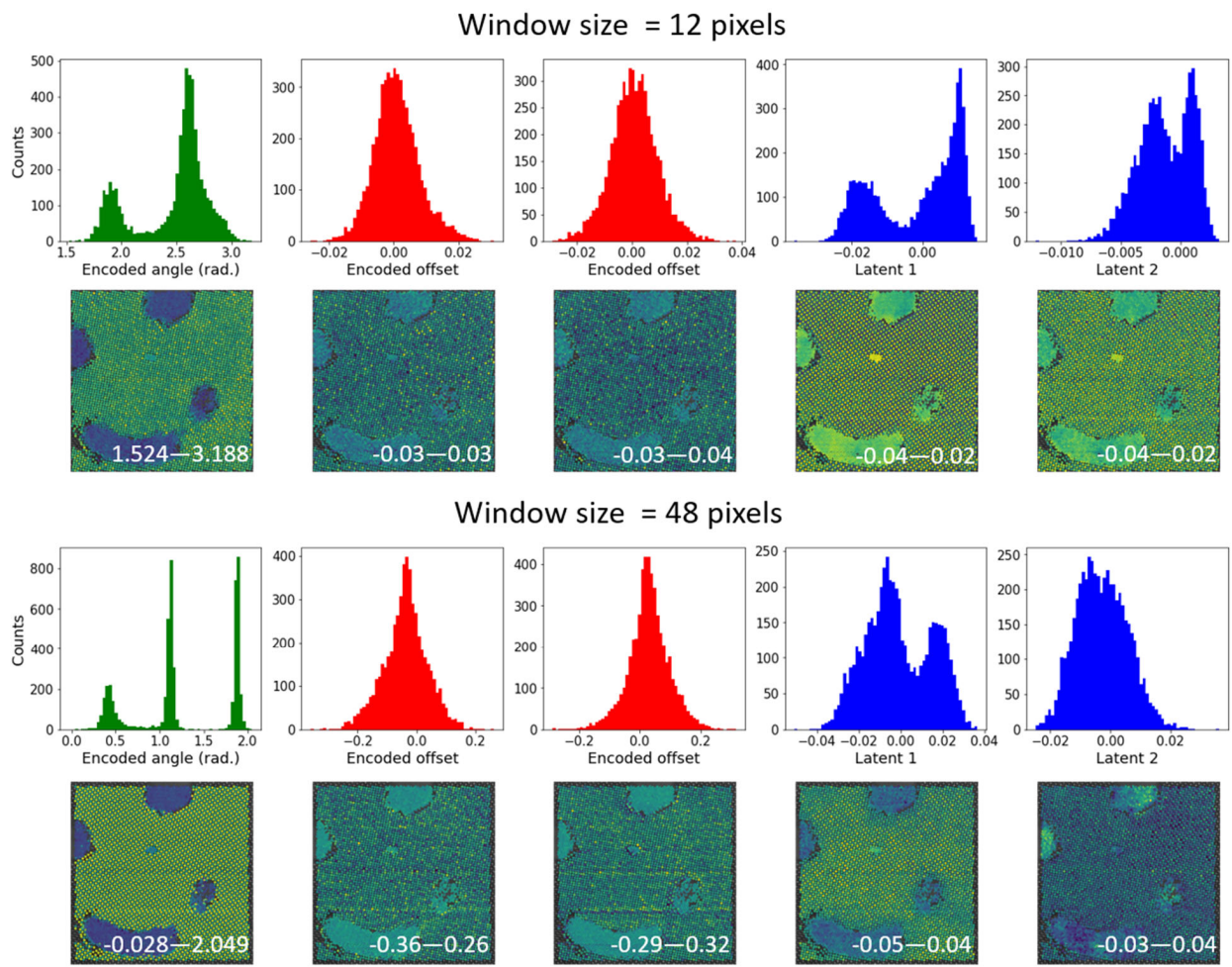

**Figure S2:** Demonstration of the effects of window size variation on the rVAE analysis for $(La_xSr_{1-x})MnO_3 - NiO$ system. Insets show range of intensity variation.

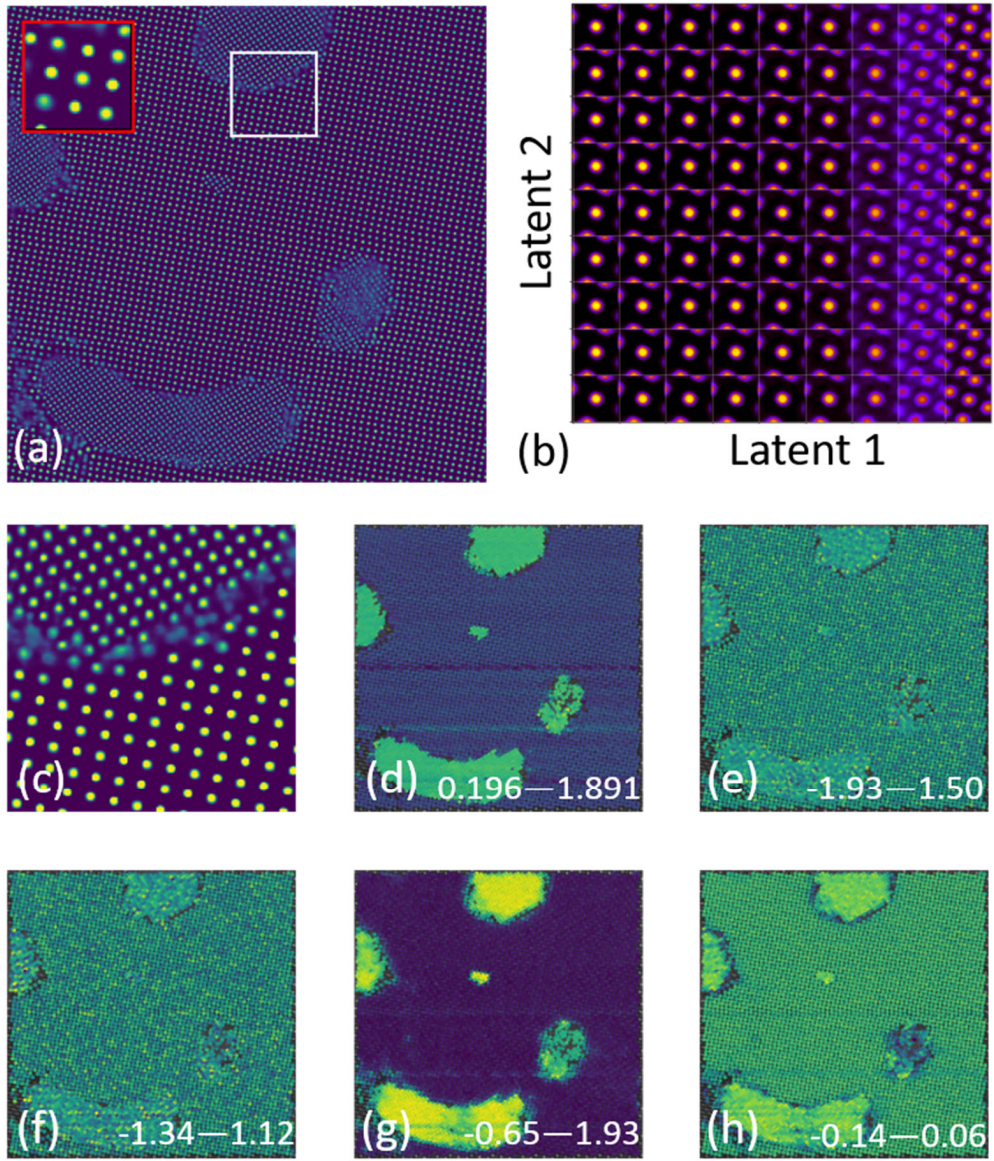

**Figure S3.** rVAE analysis of multiphase (La$_x$Sr$_{1-x}$)MnO$_3$ – NiO system using a window size of 24 pixels and DCNN segmented images. Shown are (a) segmented image, (b) sub-image representation in 2D latent parameter space, (c) zoom-in of image in (a) showing phase morphology, (d) angle, (e,f) $x$ and $y$ offsets, and (g,h) latent variables 1 and 2. Insets in (d)-(f) show range of variation in intensities.

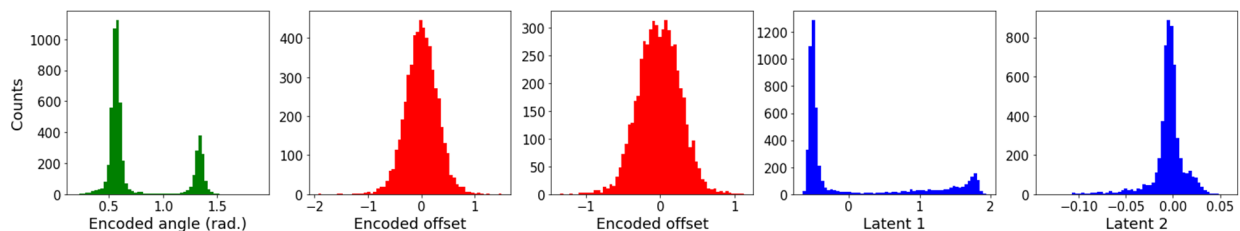

**Figure S4:** Histograms of encoded angle, offsets, and latent spaces for DCNN segmented $(La_xSr_{1-x})MnO_3 - NiO$ data shown in Fig. S3 using a window size of 24.

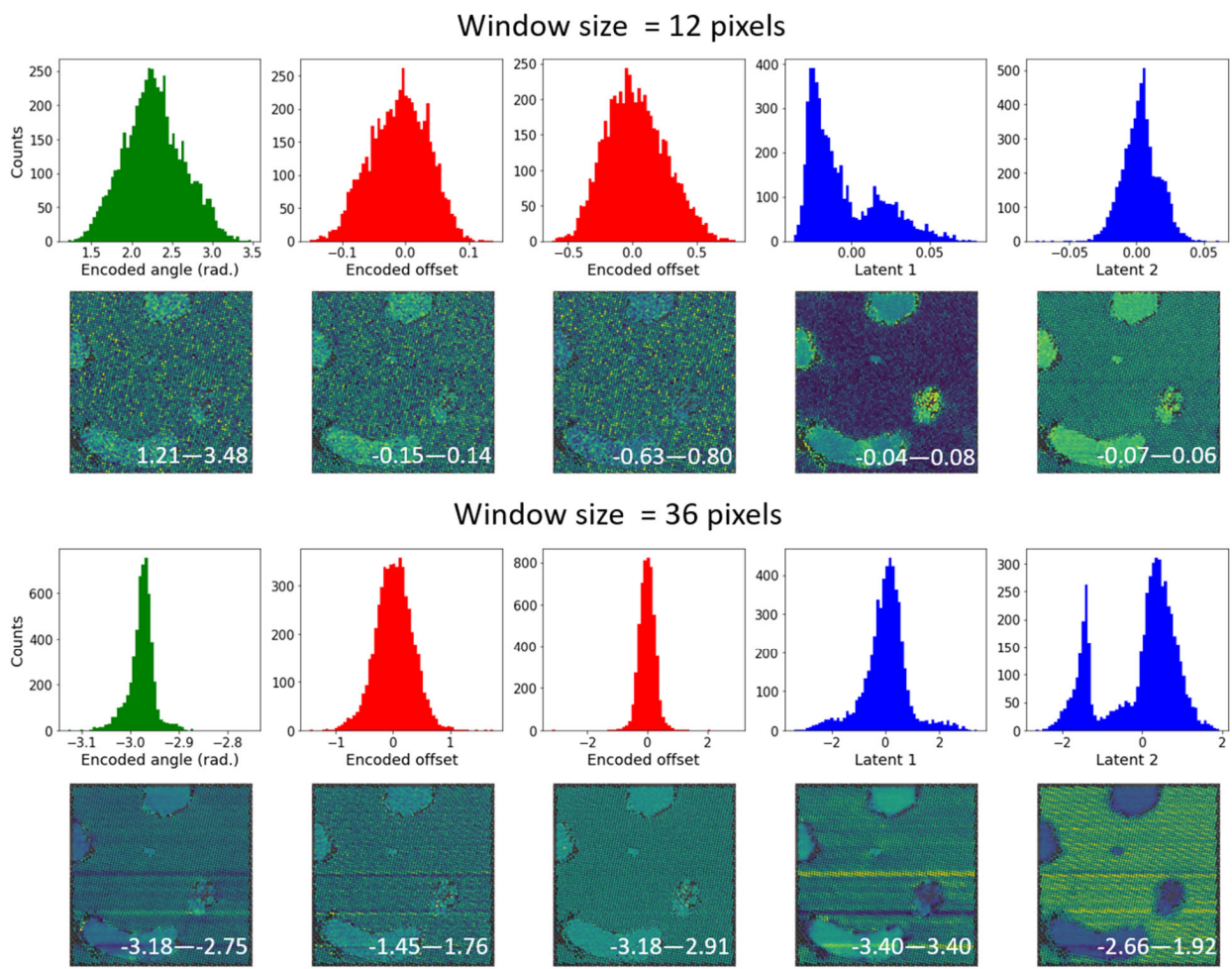

**Figure S5:** Demonstration of the effects of window size variation on rVAE analysis for DCNN segmented $(La_xSr_{1-x})MnO_3 - NiO$ data. Insets show range of intensity variation.

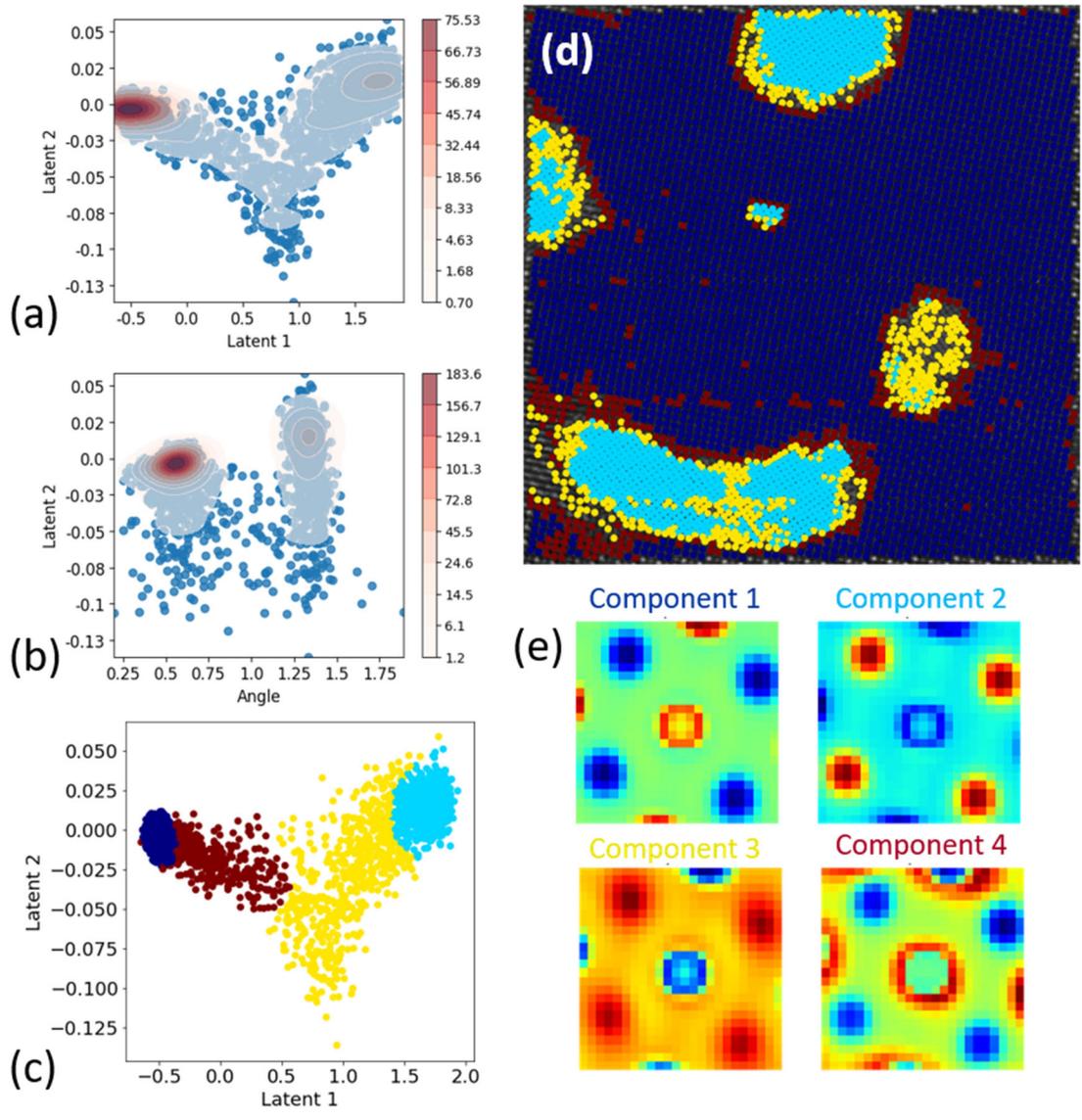

**Figure S6:** Latent space distributions. Pair distribution for (a) latent variables and (b) encoded angle and second latent variables (c) GMM clustering for latent spaces. (d) Original STEM image with superimposed class labels and GMM centroid images corresponding to four components used. (e) Four components: component 1 (dark blue), component 2 (cyan), component 3 (yellow), and component 4 (brown). Analysis performed on DCNN segmented data with a window size of 24 pixels.

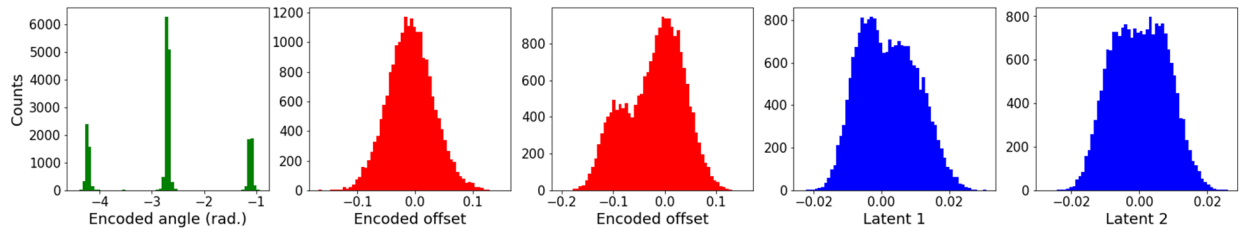

**Figure S7:** Histograms of encoded angle, offsets, and latent spaces for the $Sr_3Fe_2O_7$ system shown in Fig. 7 using a window size of 32 pixels.

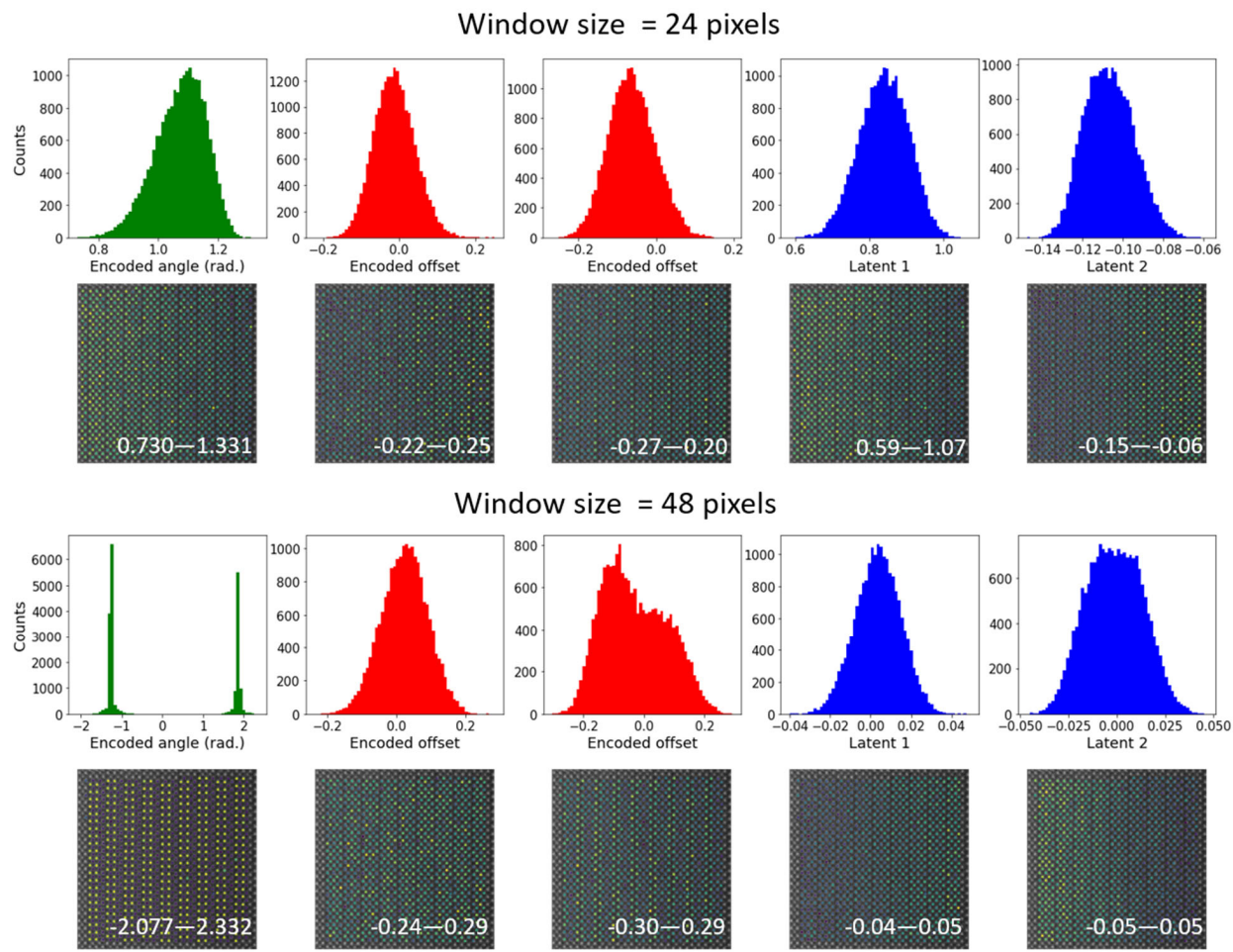

**Figure S8:** Demonstration of the effects of window size variation on rVAE analysis for $Sr_3Fe_2O_7$ system. Insets show range of intensity variations.

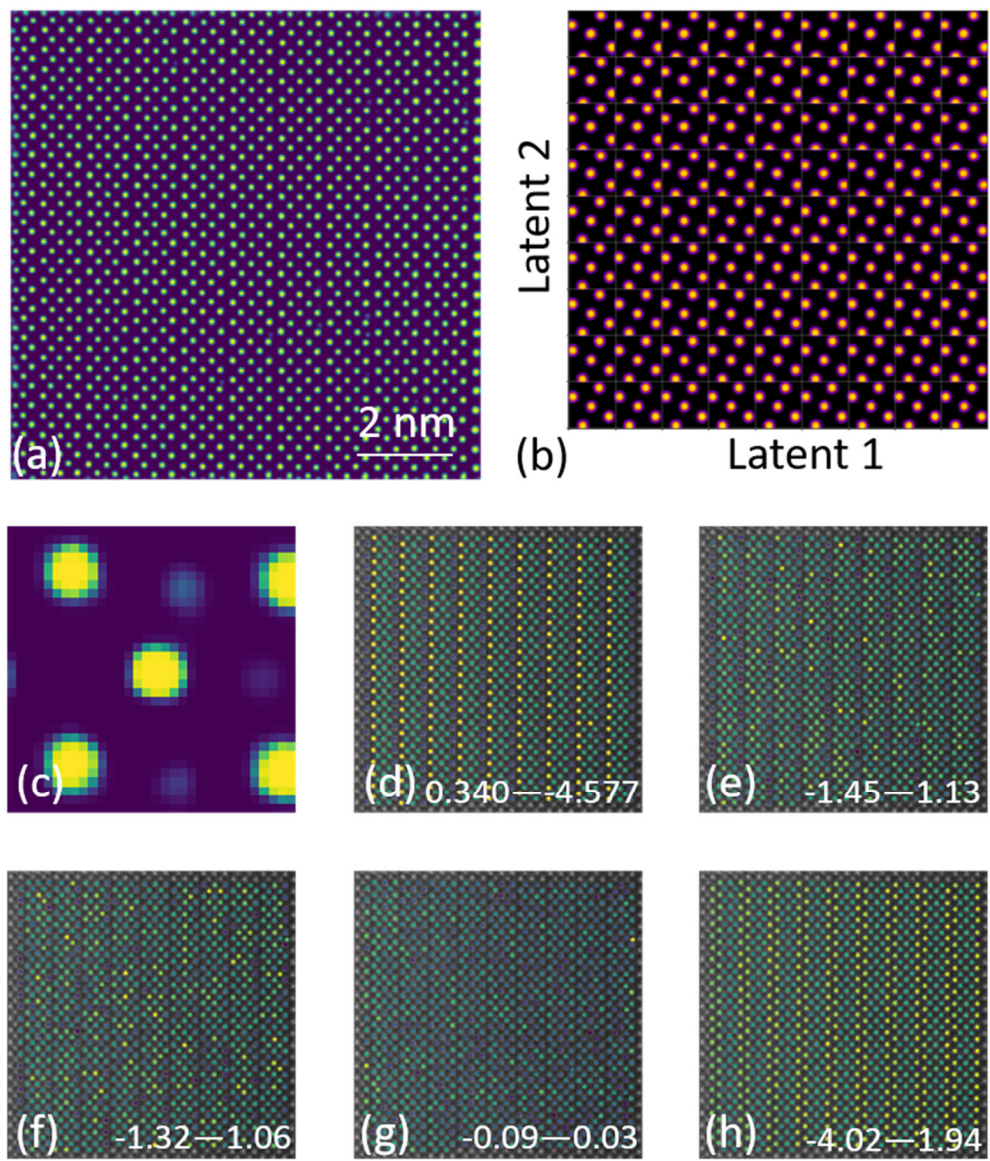

**Figure S9:** rVAE analysis of DCNN segmented $Sr_3Fe_2O_7$ image. Shown are (a) original image, (b) sub-image representation in 2D latent parameter space, (c) single sub-image used for analysis, (d) angle, (e,f) $x$ and $y$ offsets, and (g,h) latent variables 1 and 2. A window size of 32 pixels used. Insets indicate range of intensity variations in each panel.

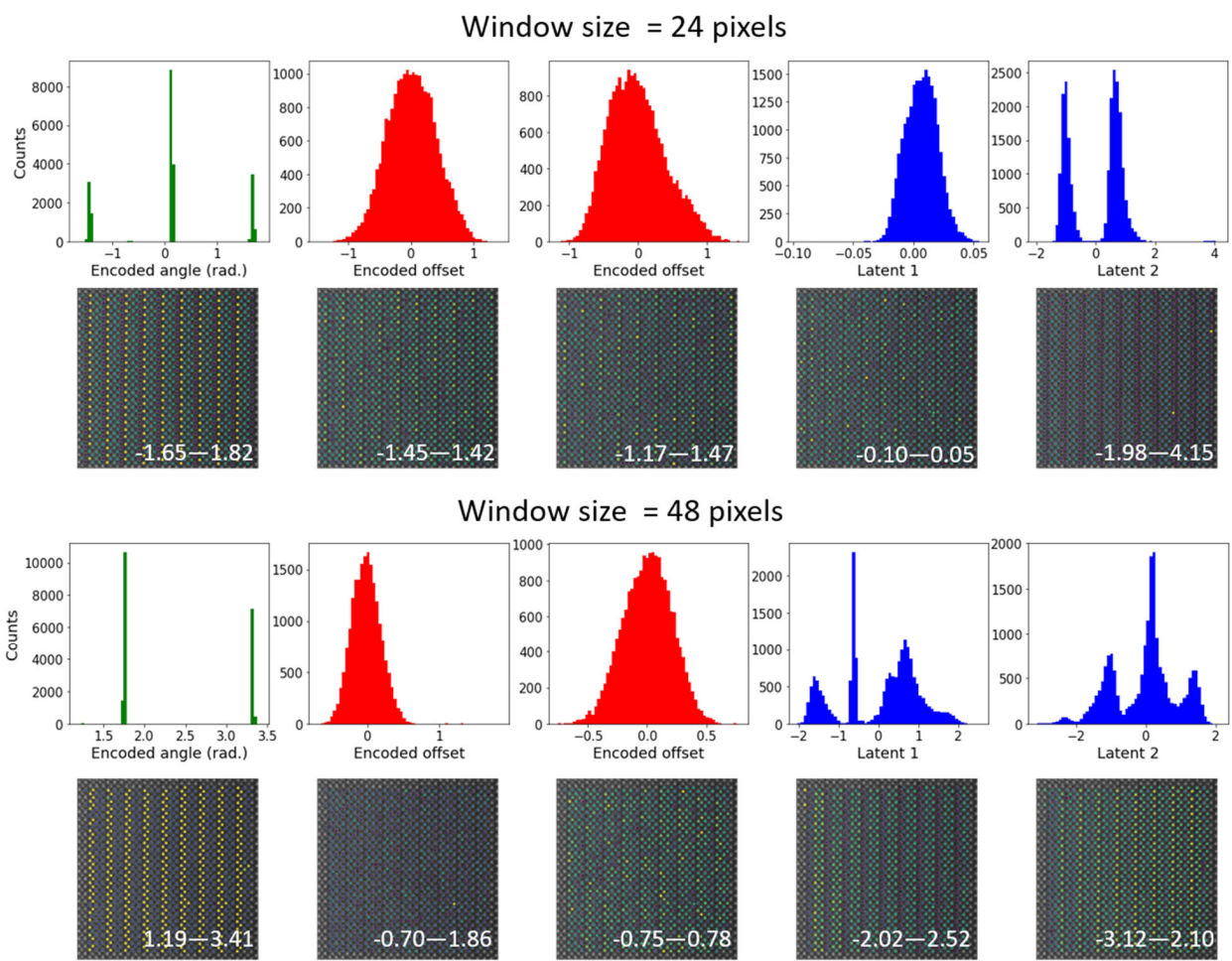

**Figure S10:** Demonstration of effects of window size variation on rVAE analysis for DCNN segmented $Sr_3Fe_2O_7$ data. Insets show range of intensity variation s.

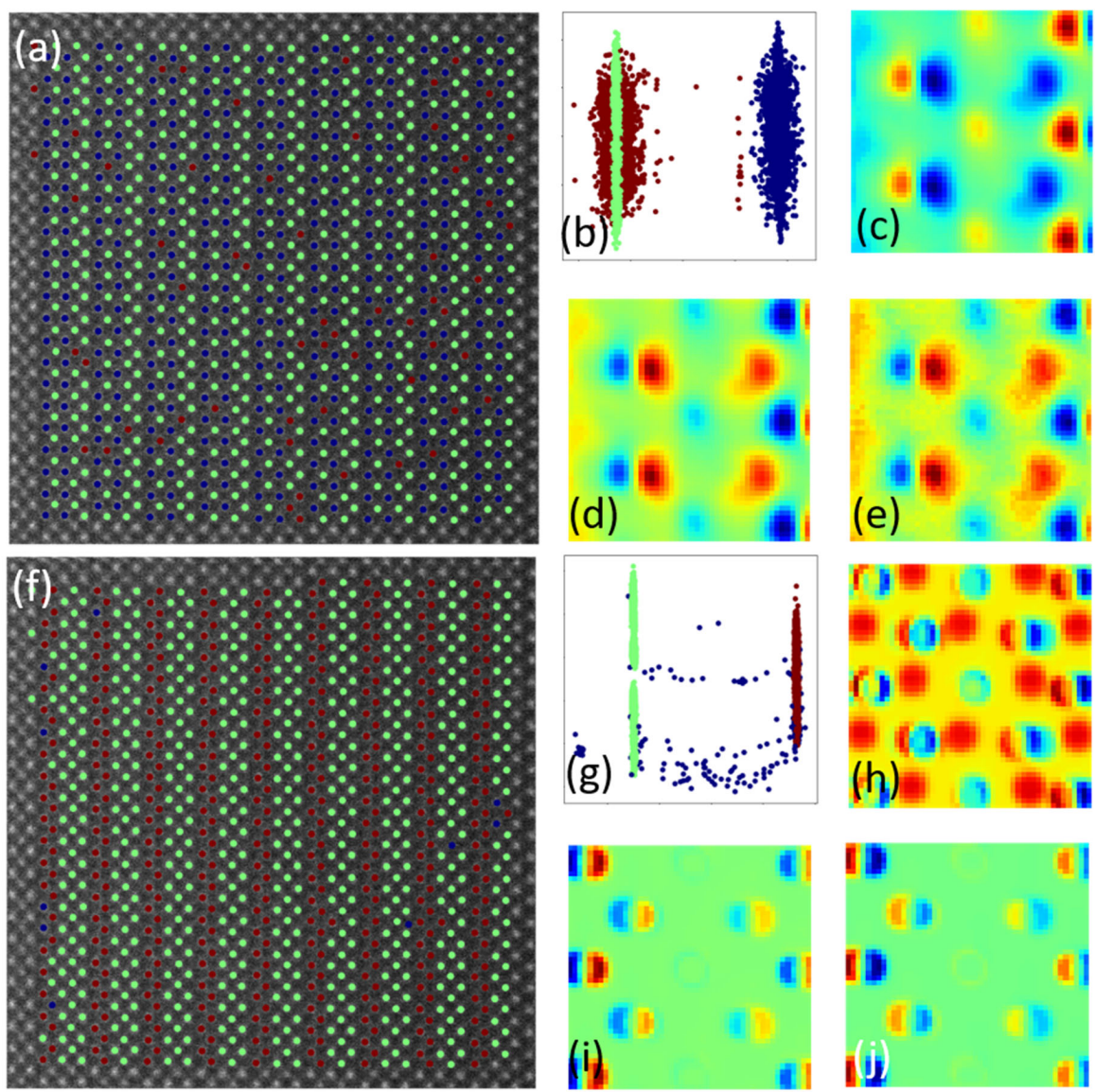

**Figure S11:** Analysis of SFO in original (a-e) and DCNN (f-j) representations. Shown are (a,f) images with superimposed GMM markers, (b,g) GMM clusters in the latent spaces, and (c-e) and (h-j) reconstructed components corresponding to the GMM centroids. Component 1 is represented by dark blue dots, with components 2 and 3 represented by green and brown dots, respectively. A window size of 48 pixels was used in both cases.

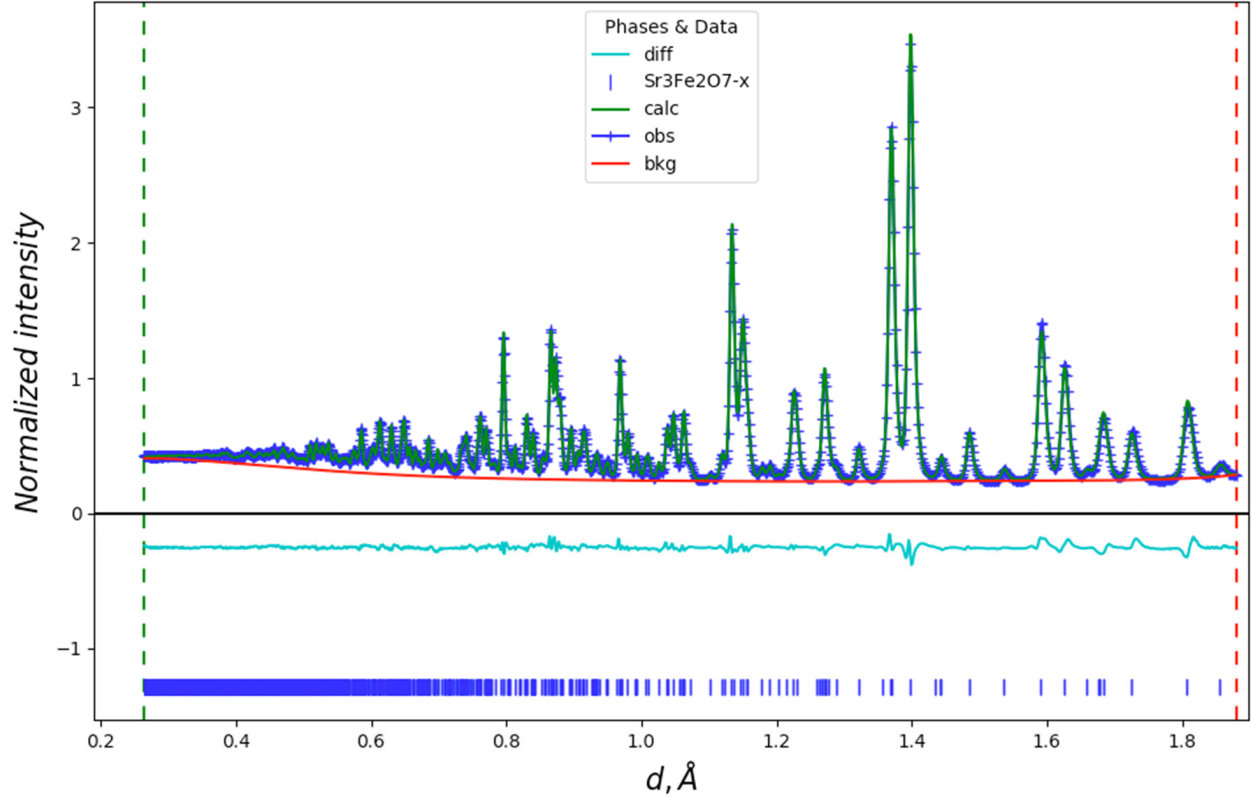

**Figure S12:** neutron powder diffraction data and refinement obtain at the NOMAD, SNS, spectrometer on powderized single crystals of $Sr_3Fe_2O_{7-x}$ at 500 K. Data was refinemed with the GSAS-II software. Data residual: $w_R$=2.7%; $R_F$=3.3% on 2489 reflections. Refined quantities are in Table S1.

**Table S1:** Refined quantities for $Sr_3Fe_2O_{7-x}$ at 500 K. Space group: I 4/m m m. Lattice parameters: a=3.87098 Å, c=20.20233 Å. Oxygen deficiency: $x$=0.2.

| Atom | x | y | z | frac | site sym | mult | $U_{iso}/Å^2$ |
|------|------|------|--------|------|----------|------|---------------|
| Sr | 0.00 | 0.00 | 0.50 | 1 | 4/mmm | 2 | 0.0095 |
| Sr | 0.00 | 0.00 | 0.3176 | 1 | 4mm | 4 | 0.0075 |
| Fe | 0.00 | 0.00 | 0.0986 | 1 | 4mm | 4 | 0.0044 |
| O | 0.00 | 0.50 | 0.0927 | 1 | mm2 | 8 | 0.0113 |
| O | 0.00 | 0.00 | 0.1941 | 1 | 4mm | 4 | 0.0120 |
| O | 0.00 | 0.00 | 0.00 | 0.78 | 4/mmm | 2 | 0.0117 |